\documentclass[superscriptaddress,twocolumn,showpacs,nofootinbib,longbibliography,notitlepage,10pt]{revtex4-2}
\usepackage{etex}
\usepackage{pgfplots}
\usepackage{physics}
\usepackage{multirow}
\usepackage{array}
\newcolumntype{C}[1]{>{\centering\arraybackslash}p{#1}}

\usepackage{amsmath,amssymb,amsthm}
\usepackage[colorlinks=true,citecolor=blue,urlcolor=blue]{hyperref}
\usepackage{times,txfonts}
\usepackage{braket}
\usepackage{color}
\usepackage{natbib}
\usepackage{amsmath,blkarray}
\usepackage{mathtools}
\usepackage{latexsym}
\usepackage{tabularx, booktabs}
\usepackage{graphics,epstopdf}
\usepackage{graphicx}
\usepackage{float}
\usepackage{graphicx}
\usepackage{amsfonts}
\usepackage{subcaption}
\usepackage{color,soul}

\begin{document}
	\title{Device-independent randomness certification using multiple copies of entangled states }
	\author{Shyam Sundar Mahato}
	\email{shyamsundarmahato59@gmail.com}
	\affiliation{National Institute of Technology Patna, Ashok Rajpath, Patna, Bihar 800005, India}
	\author{ A. K. Pan }
	\email{akp@phy.iith.ac.in}
	\affiliation{National Institute of Technology Patna, Ashok Rajpath, Patna, Bihar 800005, India}
	\affiliation{Department of Physics, Indian Institute of Technology Hyderabad, Telengana-502284, India }

	\begin{abstract}
		We demonstrate to what extent many copies of maximally entangled two-qubit states enable for generating a greater amount of certified randomness than that can be certified from a single copy. Although it appears that greater the dimension of the system implies a higher amount of randomness, the non-triviality lies in the device-independent simultaneous certification of generated randomness from many copies of  entangled states. This is because, most of the two-outcome Bell inequalities (viz., Clauser-Horne-Shimony-Holt, Elegant, or Chain Bell inequality) are optimized for a single copy of  two-qubit entangled state. Thus, such Bell inequalities can certify neither many copies of entangled states nor a higher amount of randomness. In this work, we suitably invoke a family of $n$-settings Bell inequalities which is optimized for $\lfloor n/2 \rfloor$ copies of maximally entangled two-qubit states, thereby, possess  the ability to certify more randomness from many copies of two-qubit entangled state.
	\end{abstract}

	\maketitle

	\section{Introduction} \label{SecI}
	
	Randomness is an essential resource \cite{Pironio2007, Pironio2010, Colbeck2011, Gallego2013, Silleras2014, Bancal2014, Bierhorst2018, Liu2021} for various information processing tasks such as secure communication, key distribution, encryption schemes in cryptographic protocols, statistical sampling and many others. In this regard, the generation of certifiable as well as provably-secure generation of the random number is crucial. The existing and widely used random number generators (RNGs) such as pseudo-RNG, and true-RNG lack the certification of the generated randomness \cite{Bera2017}. This is because pseudo-RNG uses a given algorithm or arithmetic operation to produce a sequence of random numbers, and thus, the privacy is based on assumptions about the computational power of the adversary as well as on the assumptions that the adversary does not know the underlying arithmetic operations or the algorithms used to produce the random numbers - making the generated pseudo-random sequence to vulnerable to adversarial guessing attack. True-RNG relies on the complexity of the calculations or the lack of complete knowledge about the systems and exploits hard-to-predict physical phenomena for generating randomness. However, in principle, one can always predict the outcomes of such physical phenomena with certainty using classical mechanics and powerful computational power \cite{Acin2016}.
	
	Now, in quantum theory, the outcomes of a measurement performed on a quantum mechanical system are truly random. This randomness persists even if we have the full knowledge of the preparation of the state of the system. Thus, such randomness does not rely on the lack of knowledge about the systems or complexity of the calculations \cite{Acin2016}. However, the fundamental problem with the certification of such randomness is that we have to rely on the inner workings of the quantum device. Even if we assume that the quantum-RNG device \cite{collantes2017} is provided by a trusted supplier, it may deviate from its intended design because of unavoidable imperfections, aging, or tampering leading to uncontrollable biases \cite{Pironio2018}. Therefore, to generate provably secure randomness, we have to certify the generated randomness in the device-independent (DI) way. 
	
	To this end, it is the empirical violation of the Bell inequality, together with the validity of the condition of signal locality, that guarantees unpredictability of Bell violating measurement outcomes, independent of the computational power of any adversary \cite{Masanes2006, Cavalcanti2012, ss2020}. Since the violation of Bell inequality only requires the input-output statistics of an experiment, the certification of the generated randomness does not rely on the inner working of the devices used. In \cite{Pironio2010}, it has been shown that the device-independent (DI) guaranteed bound of randomness is monotonic with violations of Clauser-Horne-Shimony-Holt (CHSH) \cite{CHSH1970} inequality. Further, by invoking the CHSH inequality, Colbeck \emph{et al.}\cite{Colbeck2011} introduced a protocol in which a short and weak random string can be expanded into a longer provably private random string - known as randomness expansion. Further, Acin \emph {et al.} \cite{Acin2012} shade more light on the relationship between randomness, nonlocality, and entanglement by showing more randomness can be produced from a small amount of nonlocality and states with arbitrarily little entanglement. Recently, by invoking the Hardy \cite{Hardy1992} and Cabello-Liang \cite{Cabello2002, Liang2005} relations for nonlocality, it has been shown \cite{ss2020} that the quantitative relationship between the randomness and the amount of nonlocality is more nuanced than that thought of. While the minimum guaranteed amount of randomness has a monotonic relationship with nonlocality, the maximum amount bound of randomness is quantitatively incommensurate with nonlocality.   
	
	It is important to note here that in the CHSH case, the maximum amount of guaranteed randomness that can be certified is 1.2284 bits \cite{Pironio2010} corresponding to the maximum violation of the CHSH inequality $(2\sqrt{2})$. The amount of certified randomness in the same experimental setup can not be increased even if one increases the dimension of the shared maximally entangled state. This is because the value of the CHSH function becomes optimized for the two-qubit system \cite{Cirelson1980, Supic2020, Pan2020}. Hence, increasing the dimension of the system does not provide any advantage in the generation of certified randomness. Now, it is the purpose of this manuscript to investigate the question of whether more than one copy of an entangled two-qubit state provides any advantage over a single copy for the generation of certified randomness. While at first glance, the question may seem to be obvious, deeper looks into it provide the deep-seated non-triviality. Of course, one may argue that pair of maximally entangled two-qubit states always provide a greater amount of randomness than that can be obtained from a single copy of it. Although it is trivial that pair of maximally entangled states indeed provide the quantitative advantage in the certified randomness, the non-trivial part is the certification of such a pair of entangled states. This is because, the standard CHSH inequality, as mentioned, does not provide the required certification since it can be shown that the quantum maximum of the CHSH inequality is attained for a two-qubit entangled state. Thus, the maximum violation will not be increased even if one increases the dimension of the system or use many copies of two-qubit entangled states. The Elegant Bell inequality (EBI) \cite{Gisin2007} where one party has four measurement settings and the other has three measurement settings, the maximum value is also reached for a two-qubit system \cite{Anderson2017}. Moreover, the $n$-settings chained Bell inequality cannot also certify more than one copy of maximally entangled two-qubit state for the same reason \cite{Supic2016}. 
	
	Thus, in a single experimental setup, it is not obvious to guarantee a greater amount of provably-secure certified randomness from many copies of a maximally entangled state than that obtained from a single copy. Against this backdrop, by invoking a family of many settings Bell inequality, we demonstrate that increasing the number of maximally entangled two-qubit states does provide an advantage in the generation of certified randomness. Such a family of inequalities has earlier been introduced as a dimension witness \cite{Pan2020} in the context of random-access-code communication games.
	
	This paper is organized as follows. In Sec. \ref{SecII}, we briefly recapitulate the essence of the derivation for the optimal quantum bound of the Bell inequality without assuming the dimension of the state as well as measurement (Sec. \ref{sosnop}). Next, we employ the Bell inequality through DI certification of randomness. Then, in Sec. \ref{SecIII} by suitably quantifying, we evaluate the amount of certified randomness corresponding to the optimal quantum violation of Bell inequality for an arbitrary number of measurement settings ($n$). In particular, in Secs. \ref{cr2}-\ref{cr4}, we explicitly evaluate the certified randomness when Alice and Bob share a single copy as well as more than one copy of maximally entangled two-qubit state for $n\in \{2,3,4,5,6\}$. The obtained results are illustrated in Table \ref{table1} and Fig. \ref{figrn}. Finally, we conclude our work and discussed our proposal from two different viewpoints where more than a single copy of a two-qubit maximally entangled state provides an advantage over a single copy (Sec. \ref{SecIV}).

	\section{A family of Bell inequalities and corresponding optimal quantum violations}\label{SecII}
	
	\begin{figure}[ht]
		\centering 
		{{\includegraphics[width=0.9\linewidth]{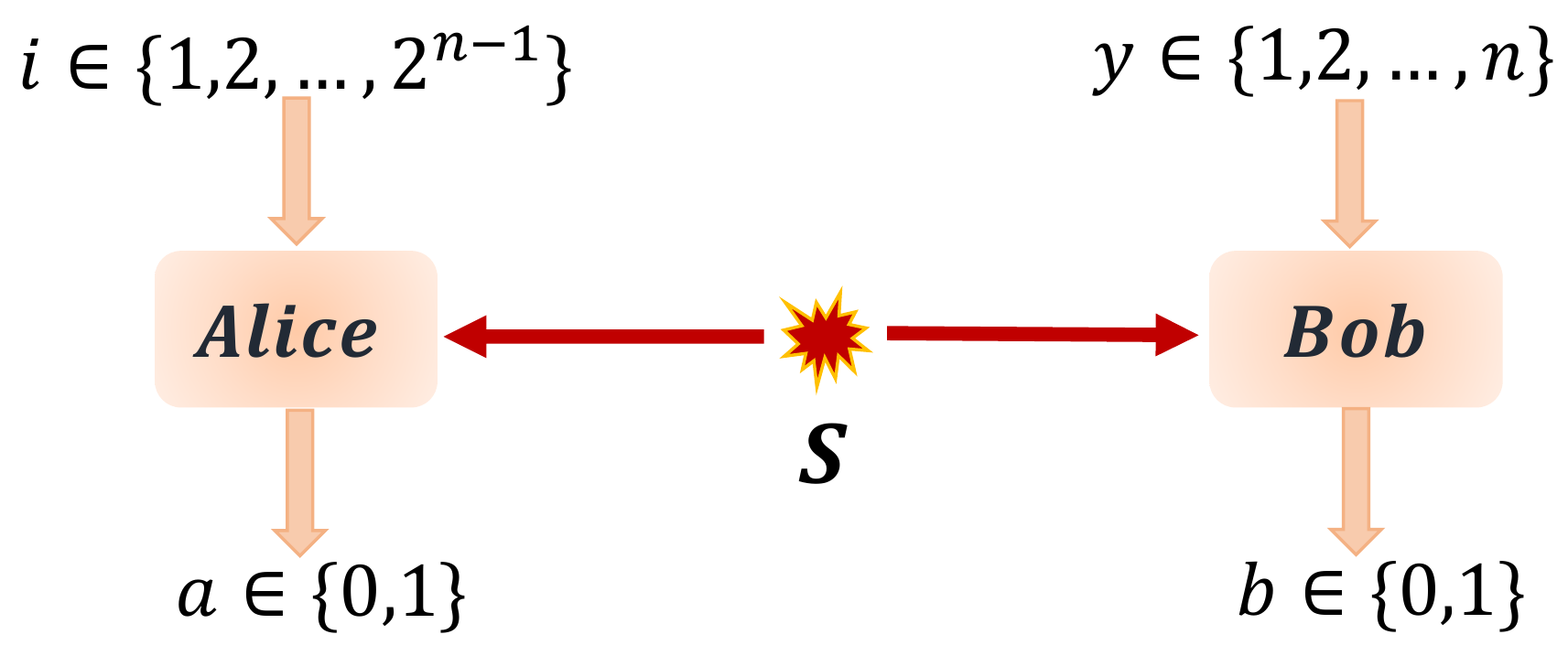}}}
		\caption{A bipartite Bell experiment involving $2^{n-1}$ settings for Alice and $n$ settings for Bob. }
		\label{bs}
	\end{figure}
	For the sake of completeness, we briefly revisit the quantum optimization of a family of Bell inequalities, which was earlier introduced as a consequence of the analysis of the $n$-bit RAC communication game \cite{Ghorai2018}. It was shown that the success probability crucially depends on the quantum violation of such Bell inequality. For our purpose of generation of certified randomness, we independently invoke this inequality. Before we begin, let us first introduce the scenario in the following.  
	
	We consider a bipartite Bell experiment where two space-like separated parties Alice and Bob perform measurements on their local subsystems. Alice performs one of $2^{n-1}$ dichotomic measurements denoted by $A_{n,i} \in \{A_{n,1}, A_{n,2}, ..., A_{n,2^{n-1}}\}$ and the corresponding outcomes be denoted as $a \in \{0,1\}$. Bob performs one of $n$ dichotomic measurements denoted by $B_{n,y} \in \{B_{n,1}, B_{n,2}, ..., B_{n,n}\}$ and the corresponding outcomes be denoted as $b\in \{0,1\}$. In the above, the following Bell functional has been proposed \cite{Ghorai2018} is given by
	\begin{align}\label{nbell}
		\mathcal{B}_{n} = \sum_{y=1}^{n}\qty(\sum_{i=1}^{2^{n-1}} (-1)^{x^i_y}  A_{n,i})\otimes B_{n,y}
	\end{align}
	The term $x^i_{y}$ takes a value of either 0 or 1 which is fixed by using the encoding scheme demonstrated in \cite{Ghorai2018, Pan2020, Pan2021}. Then for a given $i$, $x^{i}_{y}$ fix the values of $(-1)^{x^{i}_{y}}$ in the following way. Let us consider a random variable $x^{\alpha}\in \{0,1\}^{n}$ with $\alpha\in \{1,2,\cdots,2^{n}\}$. Each element of the bit string can be written as $x^{\alpha}=x^{\alpha}_{y=1} x^{\alpha}_{y=2} x^{\alpha}_{y=3},\cdots,x^{\alpha}_{y=n}$. For  example, if $x^{\alpha} = 110...10$ then $x^{{\alpha}}_{y=1} =1$, $x^{{\alpha}}_{y=2} =1$, $x^{{\alpha}}_{y=3} =0,\cdots ,x^{{\alpha}}_{y=n-1} =1,x^{{\alpha}}_{y=n} =0$. We  denote  the $n$-bit binary strings as $x^{i}$. Here we consider the bit strings such that for any two $i$ and $j$,  $x^{i}\oplus_{2} x^{j}=11\cdots1$. Clearly, we have $i\in \{1,2,\cdots,2^{n-1}\}$ constituting the inputs for Alice. If $i=1$, we get all the first bit of each  bit  string $x_y$ for every, $y\in \{1,2, \cdots ,n\}$ which are the inputs of Bob.
	
	It has been shown \cite{Pan2020} that if Alice and Bob are space-like separated, the local bound of the Bell functional $\mathcal{B}_{n}$ is given by
	\begin{equation}\label{local}
		\langle\mathcal{B}_{n}\rangle_{L}= n \ \binom{n-1}{\lfloor \frac{n-1}{2}\rfloor}
	\end{equation}
	where $\binom{m}{r}=\frac{m!}{r! \ (m-r)!}$ and $\lfloor m \rfloor$ is the floor function that takes as input a real number $m$, and gives as output the greatest integer less than or equal to $m$.
	
	Let a source $S$ emits an entangled two-qubit state $\rho_{AB}$.  It is important to note here that there exist quantum correlations for which $\langle\mathcal{B}_{n}\rangle_{Q}=Tr[\rho_{AB} \ 	\mathcal{B}_{n}]>\langle\mathcal{B}_{n}\rangle_{L}$. In the following, by employing an elegant sum-of-squares (SOS) method, the optimal quantum bound is derived.

	\subsection{The optimal quantum bound of the Bell functional $\mathcal{B}_{n}$ }	\label{sosnop}
	Here we derive the optimal quantum bound of the Bell functional $\mathcal{B}_{n}$ independent of the dimension of the state as well as measurement. For this purpose, we utilize the SOS technique introduced in \cite{Pan2020}. We consider a positive semi-definite operator $\gamma_{n}$, which can be expressed as $\gamma_{n}=\beta_{n} \mathbb{I}_{d}-(\mathcal{B}_{n})_{Q}$, where $\beta_{n}$ is the optimal value that can be obtained when $\langle \gamma_{n}\rangle$ is equal to zero. This can be proved by considering a set of operators $M_{n,i}$ which are linear functions of the observables $A_{n,i}$ and $B_{n,y}$ such that  
	\begin{align}\label{gamma}
		\gamma_{n}=\sum\limits_{i=1}^{2^{n-1}}  \frac{\omega_{n,i}}{2} \qty(M_{n,i})^{\dagger}\qty(M_{n,i})
	\end{align}
	where the operators $M_{n,i}$ and the quantities $\omega_{n,i}$ are given as follows 
	\begin{eqnarray}
		\label{mi}
		&&M_{n,i}=\frac{1}{\omega_{n,i}}\qty(\sum\limits_{y=1}^{n} (-1)^{x^i_y} B_{n,y})-A_{n,i} \nonumber \\
		&&\omega_{n,i}=\Big|\Big|\sum\limits_{y=1}^{n} (-1)^{x^i_y} B_{n,y}\ket{\psi}\Big|\Big|
	\end{eqnarray}
	where, $||\cdot||$ is the Frobenious norm, given by $|| \mathcal{O}\ket{\psi}||=\sqrt{\bra{\psi}\mathcal{O}^{\dagger}\mathcal{O}\ket{\psi}}$.
	
	Now, putting Eq. (\ref{mi}) into Eq. (\ref{gamma}) and by noting that $A_{n,i}^{\dagger} A_{n,i}=B_{n,y}^{\dagger} B_{n,y}=\mathbb{I}_{d} $, we obtain
	\begin{eqnarray} \label{opt1}
		\gamma_{n}&=&-(\mathcal{B}_{n})_{Q} + \sum\limits_{i=1}^{2^{n-1}}\left[ \frac{1}{2\omega_{n,i}}\left(\sum\limits_{y=1}^{n} (-1)^{x^i_y} B_{n,y}\right)^2 +  \frac{\omega_{n,i}}{2} \mathbb{I}_d\right] \nonumber \\
		&=& -(\mathcal{B}_{n})_{Q} +  \sum\limits_{i=1}^{2^{n-1}}\omega_{n,i} \  \mathbb{I}_d \ \ \ \ \text{[from Eq.~(\ref{mi})]}
	\end{eqnarray}
	
	Therefore, it follows from the above Eq.~(\ref{opt1}) that the quantum optimum value of $(\mathcal{B}_{n})_{Q}$ is attained when $\langle\gamma_{n}\rangle_{Q}= \bra{\psi} \gamma_{n} \ket{\psi} = 0$, which in turn provides 
	
	\begin{equation}\label{optbn}
		\langle \mathcal{B}_{n}\rangle_{Q}^{opt} = \max\left(\sum\limits_{i=1}^{2^{n-1}}\omega_{n,i}\right)
	\end{equation}
	
	Note that such optimal value will be obtained for the following optimization conditions
	\begin{equation}
		M_{n,i}=0 \implies A_{n,i}= \frac{1}{\omega_{n,i}}\sum\limits_{y=1}^{n} (-1)^{x^i_y} B_{n,y}
	\end{equation}
	
	Now, in order to achieve the optimal quantum value $\langle \mathcal{B}_{n}\rangle_{Q}^{opt}$ from Eq.~(\ref{optbn}), we first evaluate $(\sum\limits_{i=1}^{2^{n-1}}\omega_{n,i})$ by invoking the convex inequality as follows
	\begin{align}
		\label{concav}
		\sum\limits_{i=1}^{2^{n-1}}\omega_{n,i}\leq \sqrt{2^{n-1} \sum\limits_{i=1}^{2^{n-1}} (\omega_{n,i})^{2}}
	\end{align}
	
	It is crucial to remark here that in Eq. (\ref{concav}), the equality holds only when all $\omega_{n,i}$ are equal for each $i$. 
	
	Since $B_{n,y}$s are dichotomic, the quantity $\omega_{n,i}$ can explicitly be written from Eq.~(\ref{mi}) as
	
	\begin{eqnarray}\label{omega}
		\omega_{n,i}&=& \Big[ n+ \langle \{(-1)^{x^i_1} B_{n,1}, \sum\limits_{y=2}^{n}(-1)^{x^i_y} B_{n,y}\} \rangle \nonumber\\
		&+& \langle \{(-1)^{x^i_2} B_{n,2}, \sum\limits_{y=3}^{n}(-1)^{x^i_y} B_{n,y}\}\rangle + \cdots \cdots \nonumber\\
		&+&\langle \{(-1)^{x^i_{n-1}} B_{n,n-1}, (-1)^{x^i_n} B_{n,n}\}\rangle\Big]^{\frac{1}{2}}
	\end{eqnarray}
	where $\{ \ , \ \}$ denotes the anti-commutation relation.
	
	As already mentioned that the equality holds in Eq.~(\ref{concav}) when each of $\omega_{n,i}$ are equal. From the above Eq. (\ref{omega}) it can be shown that all $\omega_{n,i}$ are equal to each other if Bob's observables are mutually anti-commuting, i.e., $\{B_{n,y},B_{n,y^{\prime}}\}=0 \ \ \forall y,y^{\prime}\in \{1,2,\cdots,n\}$. This then, provides the maximum value of $\omega_{n,i}=\sqrt{n}$. Therefore form Eqs.(\ref{optbn}) and (\ref{concav}), it is now straightforward to obtain 
	\begin{equation}
		\label{opt}
		\langle\mathcal{B}_{n}\rangle_Q^{opt}= 2^{n-1}\sqrt{n}
	\end{equation}
	
	It is to be noted here that the optimal value will be achieved when there exist $n$ number of mutually anti-commuting observables in Bob's local Hilbert space, thereby, specifying the dimension of the global Hilbert pace, given by $d=4^{m}$ with $m=\lfloor n/2 \rfloor$, as well as the number of maximally entangled state ($m=\lfloor n/2 \rfloor$) required for achieving such optimal value in quantum theory.
	
	Thus, for the cases of $n\geq 4$, a single copy of a maximally entangled two-qubit state will not suffice the purpose and one requires a higher dimensional system. For example, the optimal value of the Bell expression for $n=4$ requires at least two copies of a bipartite maximally entangled qubit state.

	\section{ Generation of certified Randomness from the new family of Bell inequalities} \label{SecIII}
	
	It follows from the argument presented in the preceding Sec. (\ref{SecII}) that in the quantum theory there exist correlations for which it is possible to obtain $\langle\mathcal{B}_{n}\rangle_{L}<\langle\mathcal{B}_{n}\rangle_{Q}\leq 2^{n-1}\sqrt{n}$. This implies that such Bell violating quantum correlations cannot be reproduced by any predetermined local strategy between Alice and Bob. Thus, empirical violations of the proposed Bell inequalities certify the presence of inherent randomness in the observed statistics. It is the purpose of this manuscript to appropriately quantify such inherent certified randomness.  
	
	In information theory, among the different measures of entropy, the quantity min.-Entropy ($H_{\infty}$) \cite{Renner2009} is used in all the studies for the quantification of randomness. Here, to compare with those earlier relevant works, we also take min.-Entropy as the suitable quantifier of certified randomness. Note that the min.-Entropy is determined only by the event occurring with the highest probability, independent of the probabilities of the other events, it quantifies the minimum unpredictability involved in the probability distribution \cite{Renner2009}. Thus, min.-Entropy provides the most secure bound of the generated randomness. From now on, we call such bound as guaranteed bound of randomness. 
	
	We quantify the amount of DI certified global randomness $\qty(R_{min})_n$ for a given Bell violation, $\langle\mathcal{B}_n(\vec{\mathbb{P}}_{obs}^{n})\rangle> n \ \binom{n-1}{\lfloor \frac{n-1}{2}\rfloor}$, in terms of the min.-Entropy as follows 
	\begin{eqnarray} \label{randef}
		\qty(R_{min})_n &=& \min\limits_{\vec{\mathbb{P}}_{obs}} H_{\infty}(a,b|A_{n,i},B_{n,y}) = -\log_2\qty[\max_{\vec{\mathbb{P}}_{obs}} p(a,b|A_{n,i},B_{n,y})]  \nonumber\\
		&& \ \ \ \text{subject to:} \nonumber\\
		&&  \ \ \ \text{(i)} \ \vec{\mathbb{P}}_{obs} \in \{p(a,b|A_{n,i},B_{n,y})\},\nonumber \\ 
		&&  \ \ \ \text{(ii)} \  p(a,b|A_{n,i},B_{n,y})=Tr[\rho_{AB} \ (\Pi_{A_{n,i} }^a \otimes \Pi_{B_{n,y}}^b)],\nonumber \\
		&& \ \ \ \text{(iii)} \  \langle\mathcal{B}_{n}\rangle_Q=\left\langle\sum_{y=1}^{n}\sum_{i=1}^{2^{n-1}} (-1)^{x^i_y}  A_{n,i}\otimes B_{n,y}\right\rangle = \epsilon \nonumber\\
		&&\ \ \ \text{with} \ n \ \binom{n-1}{\lfloor \frac{n-1}{2}\rfloor}<\epsilon\leq 2^{n-1}\sqrt{n} \ ;
	\end{eqnarray}
	where $\vec{\mathbb{P}}_{obs}^{n} \in {\mathbb{R}^{n 2^{n+1}}}$ is a vector in a $(n 2^{n+1})$ dimensional real number space, which denotes the set of all observed joint probability distributions known as behaviour. 
	
	It is important to remark here that for a given Bell violation, there may exist more than one behaviour. Thus, to ensure the security of the quantified randomness, for a given Bell violation, the min.-Entropy need to be minimized over all possible behaviours. 
	
	In this regard, a point to be noted here is that amount of certified randomness can be quantified in two different ways \cite{ss2020}. To understand it more clearly, let us first consider the simplest 2 parties-2 measurements per party-2 outcomes per measurement (2-2-2) scenario. In such a scenario, there are four possible combinations of the pairs of measurement settings, given by $\{(A_1,B_1),(A_1,B_2),(A_2,B_1),(A_2,B_2)\}$, and one can always evaluate the maximum value of joint probability corresponding to each combination of pairs of measurement settings, denoted by $P^{\ast}_{ij} \ \ \forall i,j\in\{1,2\}$. Then, the amount of randomness corresponding to each such combination is given by $R_{ij}=-\log_2\qty[P^{\ast}_{ij}]$. Now, we can evaluate the amount of certified randomness in two ways - (i) by taking the minimum value of $R_{ij}$ which then gives us the DI guaranteed bound of randomness, $(R_{min}=\min\limits_{i,j} R_{ij})$, and (ii) by taking the maximum value of $R_{ij}$ which then gives us the maximum amount of randomness, $(R_{max}=\max\limits_{i,j} R_{ij})$. Such a discussion on the quantification of the certified amount of randomness is essential since there are works \cite{Acin2012, Law2014, ss2020} that have used $R_{max}$ as a quantifier of certified randomness to show that close to two bits of randomness can be certified in the 2-2-2 scenario. Further, by invoking a suitable generalised measurement (POVM) scheme, it has been shown \cite{Anderson2018, Woodhead2019} that it is possible to achieve close to four bits of amount of $R_{max}$ certified by the Elegant Bell inequality.
	
	Importantly, in our prescription of quantifying the DI certified randomness, we take $R_{min}$ as a viable quantifier of the certified randomness as given by Eq.~(\ref{randef}) like the bound considered for the estimation of DI certified randomness by Pironio \emph{et al.} \cite{Pironio2010}. Moreover, a significant point to be reflected here is that for the optimal violation of the concerned Bell functional ($\mathcal{B}_n$), all the maximum joint probabilities corresponding to each combination of pairs of measurement settings are found to be equal, thereby in our treatment $R_{min}=R_{max}$. However, while one can also interject a suitable measurement scheme of POVM and/or employ higher settings tilted Bell inequalities to certify a greater amount of randomness, such a line of study is beyond the scope of our present manuscript. 
	
	Now, since the quantum theory is not a polytope, it is impossible to analytically evaluate the guaranteed bound of randomness by executing the optimization by taking into account all possible behaviours. However, for the optimal quantum violation, the observed behaviour is unique. Then, for optimal quantum violation of a Bell inequality, it will be straightforward to evaluate the secure guaranteed bound of randomness. Hence, for our purpose of the evaluation of the certified randomness in the quantum theory, we take recourse to the case when the proposed Bell inequality is optimally violated. 
	
	Before we proceed to evaluate the amount of certified randomness, we again point out the interesting feature anew: by increasing the dimension of the system, whether it is possible to certify a greater amount of guaranteed randomness than that obtained in the CHSH case in a same experimental setup. Note that such an advantage cannot be revealed by using only the CHSH inequality, EBI, and Chain Bell inequality. The deep-seated significance of the proposed newly found family of Bell inequalities is that the maximal quantum violation provides a dimension witness \cite{Pan2020}. Thus, many copies of maximally entangled two-qubit states can be certified from the optimum quantum violation of such inequality by increasing the number of measurement settings. This, then, will provide the advantage in the randomness generation in a provable-secure way than the CHSH or elegant or chain Bell inequalities. We are now in a position for evaluating the certified randomness corresponding to the optimal quantum violations of the Bell inequalities given by $\mathcal{B}_n$ for different values of $n$.

	\subsection{Evaluation of certified randomness for $n=2$} \label{cr2}
	
	Note that the $n=2$ case corresponds to the standard CHSH inequality. It has been shown (Sec. \ref{sosnop}) that to ensure the DI maximal violation of the CHSH inequality, both Alice's and Bob's observables need to be anti-commuting. The such maximum value is attained for the two-qubit maximally entangled state. Hence, we can always construct such anti-commuting observables in the local two-dimensional Hilbert space. Then, a straightforward algebra completely characterizes the unique behaviour $\vec{\mathbb{P}}_{obs}^2$ with the maximum joint probability $p_{2}^{\ast}$ given by
	\begin{equation}
		p_{2}^{\ast}= \frac{1}{4}(1+\frac{1}{\sqrt{2}})
	\end{equation} 
	and subsequently, the amount of randomness is given by
	\begin{equation}
		\qty(R_{min})_{n=2} = -\log_2\left[\frac{1}{4}\left(1+\frac{1}{\sqrt{2}}\right)\right] \approx 1.2284 \ bits
	\end{equation} 
	
	Note the amount of randomness for $n=2$, $ 	\qty(R_{min})_{n=2}=1.2284 \ bits$ corresponding to the optimal Bell violation (here CHSH) is found to be the same as earlier obtained in \cite{Pironio2010}.

	\subsection{Evaluation of Certified Randomness for $n=3$} \label{cr3}
	The Bell inequality $\mathcal{B}_n$ for $n=3$ corresponds to the EBI given by
	\begin{eqnarray}\label{bell3}
		\langle\mathcal{B}_{3}\rangle_{L} &=& A_{3,1} \otimes \left(B_{3,1}+B_{3,2}+B_{3,3}\right) \nonumber \\
		&+& A_{3,2} \otimes \left(B_{3,1}+B_{3,2}-B_{3,3}\right) \nonumber\\
		&+&A_{3,3} \otimes \left(B_{3,1}-B_{3,2}+B_{3,3}\right) \nonumber \\
		&+& A_{3,4} \otimes \left(-B_{3,1}+B_{3,2}+B_{3,3}\right) \leq \ 6
	\end{eqnarray}
	
	It is interesting to note here that while the EBI does not provide any quantum advantage (or the QM optimal violation) even if one increases the system's dimension such Bell inequality provides a greater amount of certified randomness than that obtained in the CHSH case.
	
	Now, since the optimal quantum violation $	\langle\mathcal{B}_3\rangle_{Q}$ of the EBI corresponds to a unique statistics which subsequently fixes the least requirement of the shared state to be the maximally entangled two-qubit state as well as puts the restriction on the observables for Both Alice and Bob. Here we construct the desired observable in the two-dimensional local Hilbert space from the conditions for quantum optimality as follows
	\begin{eqnarray}\label{obs3}
		&&A_{3,1}= (\sigma_{x} + \sigma_{y}+\sigma_{z})/{\sqrt{3}} \ ; \ A_{3,2}=(\sigma_{x} + \sigma_{y} - \sigma_{z})/{\sqrt{3}}\nonumber\\
		&& A_{3,3}= (\sigma_{x} - \sigma_{y} + \sigma_{z})/{\sqrt{3}} \ ; \ A_{3,4}=(-\sigma_{x} + \sigma_{y} + \sigma_{z})/{\sqrt{3}}\nonumber\\
		&&B_{3,1}=\sigma_{x} \ \ ; \ \ B_{3,2}=\sigma_{y} \ \ ; \ \ B_{3,3}=\sigma_{z}
	\end{eqnarray}
	
	Now, by employing the observables given in Eq. (\ref{obs3}) and the maximally entangled state $\rho_{AB}=\ket{\psi^{-}}\bra{\psi^{-}}$, where $\ket{\psi^{-}}=\frac{1}{\sqrt{2}}(\ket{01}-\ket{10})$, the observed behaviour $\vec{\mathbb{P}}_{obs}\equiv \{p(a,b|A_{3,i},B_{3,y},\rho_{AB})\}\in\mathbb{R}^{48}$ can be evaluated. We found that for the EBI, the maximum violation leads to the unique behaviour having joint probabilities of the following form
	\begin{eqnarray}\label{maxp3}
		p(a,b|A_{3,i},B_{3,y},\rho_{AB})&=& Tr[\rho_{AB} (\Pi_{A_{3,i}}\otimes\Pi_{B_{3,y}})]\nonumber\\
		&=&\frac{1}{4}(1\pm\frac{1}{\sqrt{3}})
	\end{eqnarray}
	with maximum joint probability $p_{3}^{\ast}=\frac{1}{4}(1+\frac{1}{\sqrt{3}})$ and consequently, the guaranteed amount of randomness is given by
	\begin{equation}
		\qty(R_{min})_{n=3} = -\log_2 \qty[\frac{1}{4}\qty(1+\frac{1}{\sqrt{3}})] = 1.3425 \ bits
	\end{equation}
	
	It is to be noted here that the amount of guaranteed randomness is found to be greater than that obtained in the CHSH case. Hence, for the generation of a guaranteed amount of randomness, the use of EBI will be more advantageous than that of CHSH inequality. Therefore, it seems that increasing the number of measurement settings may provide an advantage in the generation of certified randomness. Next, in the following, we proceed to evaluate the same for $n=4$.

	\subsection{Evaluation of Certified Randomness for $n=4$} \label{cr4}
	The Bell inequality $\mathcal{B}_n$ for $n=4$ is given as follows
	\begin{eqnarray}\label{bell4}
		\langle\mathcal{B}_{4}\rangle_{L}  &=& A_{4,1} \otimes\left( B_{4,1}+B_{4,2}+B_{4,3}+B_{4,4}\right) \nonumber\\
		&+& A_{4,2} \otimes\left( B_{4,1}+B_{4,2}+B_{4,3}-B_{4,4}\right)\nonumber \\
		&+& A_{4,3} \otimes\left( B_{4,1}+B_{4,2}-B_{4,3}+B_{4,4}\right)\nonumber \\
		&+& A_{4,4} \otimes\left( B_{4,1}-B_{4,2}+B_{4,3}+B_{4,4}\right)\nonumber \\
		&+& A_{4,5} \otimes\left( -B_{4,1}+B_{4,2}+B_{4,3}+B_{4,4}\right)\nonumber \\
		&+& A_{4,6} \otimes\left( B_{4,1}+B_{4,2}-B_{4,3}-B_{4,4}\right)\nonumber\\
		&+& A_{4,7} \otimes\left( B_{4,1}-B_{4,2}+B_{4,3}-B_{4,4}\right)\nonumber\\
		&+& A_{4,8} \otimes\left( B_{4,1}-B_{4,2}-B_{4,3}+B_{4,4}\right) \leq \ 12
	\end{eqnarray}
	
	It has been shown that if Alice and Bob shares a single copy of maximally entangled state, the above Bell inequality given by the Eq.~(\ref{bell4}) is optimised if Bob's observables satisfy the condition $\{B_{4,1},B_{4,2}\}=\{B_{4,1},B_{4,3}\}=\{B_{4,1},B_{4,4}\}=\{B_{4,2},B_{4,3}\}=\{B_{4,2},B_{4,4}\}=0$ and $\{B_{4,3},B_{4,4}\}=\pm2$. Then, by using such constraint relations we construct the following observables in Hilbert space dimension two $(\mathcal{H}^2)$
	\begin{eqnarray}\label{nobs4}
		&& B_{4,1} = \sigma_{x} \ ; \ \ B_{4,2} = \sigma_{y} \ ; \ \ B_{4,3} =  B_{4,4} = \sigma_{z} \nonumber \\
		&& A_{4,i}=\frac{1}{\sqrt{N_i}}\sum\limits_{y=1}^{4} (-1)^{x^i_y} B_{4,y}
	\end{eqnarray}
	where $N_i =2 \ \forall \{2,3,7,8\}$ and $N_i =6 \ \forall \{1,4,5,6\}$.
	Then one can evaluate the behaviour $\vec{\mathbb{P}}_{obs}^4\equiv \{p(a,b|A_{4,i},B_{4,y},\rho_{AB})\}\in\mathbb{R}^{128}$ corresponding to such observables given by Eq.~(\ref{obs4}) and a single copy of maximally entangled state. The maximum Bell value in this case is then given by $4(\sqrt{2}+\sqrt{6})$. The greatest joint probability $(p^{\prime}_4)^{\ast}$ in this scenario is then given by
	\begin{equation}\label{nmaxp4}
		(p^{\prime}_4)^{\ast}=\frac{1}{12} \qty(3+\sqrt{6})
	\end{equation}
	Subsequently, the amount of randomness $\qty(R^{\prime})_{n=4}$ is given by
	\begin{equation}\label{nr4}
		\qty(R^{\prime})_{n=4}=-\log_2\qty[\frac{1}{12} \qty(3+\sqrt{6})] \approx 1.1388 \ bits
	\end{equation}
	
	It is important to note that such evaluated randomness $\qty(R^{\prime})_{n=4}$ from the single copy of a maximally entangled state does not provide the required security of the bound of randomness. This is because single copy of a maximally entangled state only provides a sub-optimal violation of the concerned Bell inequality, and thus, the behaviour leading to such Bell violation is not unique. Hence, in order to evaluate the DI guaranteed bound of randomness, one has to consider all possible convex combinations of local and nonlocal behaviours that produce all behaviours leading to the same Bell violation of $4(\sqrt{2} + \sqrt{6})$. For this purpose, in order to evaluate the DI guaranteed bound of randomness corresponding to a sub-optimal Bell violation, one has to invoke the numerical method proposed in \cite{Silleras2014}. However, in our present work, although the evaluated amount of randomness evaluated does not correspond to the DI secure guaranteed bound, it serves the purpose of our work that for higher settings, many copies of the maximally entangled state provide an advantage in the generation of guaranteed amount of randomness than that can be obtained by using a single copy.
	
	The randomness obtained for the single copy of the maximally entangled two-qubit state is less than that obtained for both the CHSH and Elegant cases. This is because the optimal quantum value of $\mathcal{B}_4$ is given by $\langle\mathcal{B}_{4}\rangle_{Q}^{opt}=16$ which is greater than that obtained for a single copy of a maximally entangled state. It is crucial to remark here that although the constraint relations for Alice's observables are the same for achieving  $\langle\mathcal{B}_{4}\rangle_{Q}^{opt}=16$, the constraint relations for Bob's observable are found to be mutually anti-commuting which is different from that obtained for the single copy case. Now, since the existence of four anti-commuting observables is not possible in two-dimensional Hilbert space, at least two copies of a maximally entangled state are necessarily shared between Alice and Bob with the local Hilbert dimension to be $4$. Thus, the Bell functional $\mathcal{B}_4$ is not optimized for a single copy of a maximally entangled two-qubit state.  
	
	Now, we construct the necessary observables in the local dimension $4$ in the following.
	\begin{eqnarray}\label{obs4}
		&& B_{4,1}=\sigma_{x} \otimes \sigma_{x} \ 	; \ \ B_{4,2} = \sigma_{x} \otimes \sigma_{y} \nonumber \\
		&& B_{4,3} = \sigma_{x} \otimes \sigma_{z} \ ; \ \ B_{4,4} = \sigma_{y} \otimes \mathbb{I}_2 \nonumber \\
		&&A_{4,i} = \frac{1}{2} \sum\limits_{y=1}^{4} (-1)^{x^i_y} B_{4,y}
	\end{eqnarray}
	
	The state for which $\mathcal{B}_4$ is optimized is given by
	\begin{equation}\label{state4}
		\rho_{AB}^{\otimes 2}= \rho_{AB} \otimes \rho_{AB}
	\end{equation}
	
	In this case, the maximum joint probability ($p^{\ast}_4$) is given by
	\begin{equation}\label{pmax4}
		p^{\ast}_4 = 3/8
	\end{equation}
	which in turn gives the amount of DI certified randomness as follows
	\begin{equation}
		\qty(R_{min})_{n=4} = -\log_2 (3/8) \approx 1.4150 \ bits
	\end{equation}
	
	Note that the amount of randomness for two copies of the maximally entangled two-qubit state is significantly increased than that obtained for the single copy of the maximally entangled two-qubit state. Thus, it is revealed that more than one copy of the bipartite maximally entangled state provides an advantage in the generation of certified randomness by using such Bell inequality. 
	
	We further extend our study for $n=5$ and $n=6$ cases to demonstrate how an increase in the number of shared maximally entangled two-qubit states provides more randomness over a lesser number of bipartite maximally entangled states. For $n=5$, the Bell functional $\mathcal{B}_5$ is also optimized for a pair of maximally entangled two-qubit states like for the case of $n=4$. In this case, the amount of randomness for the single copy of two-qubit entangled state is $(2-\log_2[1+\frac{\sqrt{2}+1}{\sqrt{2 \sqrt{2}+5}}])\approx 1.1025$ bits. Note that for a single copy, the amount of randomness is decreased than that of $n<5$.  However, for the pair of maximally entangled two-qubit state, the amount of DI certified randomness is $(2-\log_2[1+\frac{1}{\sqrt{5}}])\approx 1.4667$ bits. Hence, for the pair of maximally entangled two-qubit states, we find that the amount of DI certified randomness is increased than that of $n=4$.
	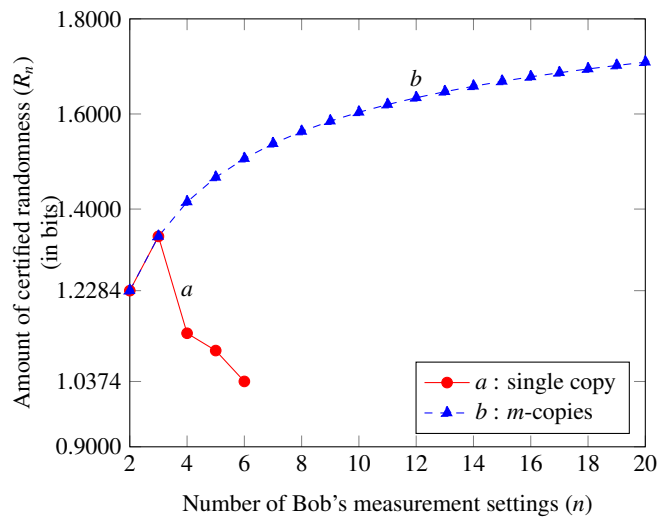
\begin{figure}[H]
		\centering
		\begin{tikzpicture}
			\begin{axis}[legend pos=south east, legend cell align=left, enlargelimits=false, xlabel={Number of Bob's measurement settings ($n$)}, ylabel style ={align=center}, ylabel= {Amount of certified randomness ($R_n)$ \\ (in bits)}, xticklabel style={ 
					/pgf/number format/fixed, /pgf/number format/fixed zerofill,
					/pgf/number format/precision=0
				}, scaled ticks=false, xtick={2,4,6,8,10,12,14,16,18,20}, yticklabel style={ 
					/pgf/number format/fixed, /pgf/number format/fixed zerofill,
					/pgf/number format/precision=4
				}, scaled ticks=false, ytick={ 0.9,1.0374, 1.2284,1.4, 1.6,1.8}, xmin=2, xmax=20, ymin=0.9, ymax=1.8
				]
				\addplot[mark=*,
				mark size=2pt,color=red]
				table[meta=R]
				{rNqubit.txt};
				\addlegendentry{ \textit{a} : single copy }
				\addplot[mark=triangle*, mark size = 2.5 pt, color = blue, dashed]
				table[meta=R]
				{rN.txt};
				\addlegendentry{ \textit{b} : $m$-copies}
				\node[above] at (20,300) {$a$};
				\node[above] at (100,740) {$b$};
			\end{axis}
		\end{tikzpicture}
		\caption{The figure illustrates the variation of the amount of certified randomness $(R_{n})$ corresponding to the optimal quantum Bell violation with different measurement settings of Bob. In particular, the red curve $`a$' shows the amount of randomness for different $n$, when a single copy of a maximally entangled two-qubit state is shared between Alice and Bob. It is found that the amount of randomness decreases with the increase of the number of measurement settings $n$. On the other hand, if $m=\lfloor n/2 \rfloor$ copies of bipartite maximally entangled state are shared between Alice and Bob, then the amount of randomness increases with the increase of $n$ as shown by the dashed blue curve $`b$'. Thus, $\lfloor n/2 \rfloor$ copies of the bipartite maximally entangled state provide an advantage over a single copy in the quantification of certified randomness.} 
		\label{figrn}
	\end{figure}
	For $n=6$, the Bell inequality $\mathcal{B}_6$ is optimized for three bipartite maximally entangled states. Interestingly, in this case, we find that the amount of DI guaranteed  randomness that can be certified is $(2-\log_2[1+\frac{1}{\sqrt{6}}])\approx1.5061$ bits when three copies of maximally entangled two-qubit state are shared between Alice and Bob. However, the amounts of  randomness for the single copy and the pair of copies are given by $(2-\log_2[1+\frac{3}{\sqrt{10}}])\approx1.0375$ and $(2-\log_2[1+\frac{1}{\sqrt{2}}])\approx 1.2284$ bits respectively. It is crucial to note here that for a single copy or pair of maximally entangled two-qubit states, there are significant decreases in the amount of randomness than that obtained for the earlier respective cases. The comparative results for all the cases are illustrated in Table~\ref{table1}.
	
	\begin{widetext}
		\begin{table*}[ht]
			\centering
			\begin{tabular}{|C{0.9in}|C{1.8in}|C{1.6in}|C{1.6in}|}
				\hline \cline{1-4}
				\multicolumn{1}{|c|}{\multirow{2}{*}{Number of Bob's}}&
				\multicolumn{3}{c|}{} \\
				\multicolumn{1}{|c|}{\multirow{2}{*}{measurement}}&
				\multicolumn{3}{c|}{Amount of certified randomness for $m\leq \lfloor n/2 \rfloor$ copies of maximally entangled two-qubit state} \\
				\multicolumn{1}{|c|}{\multirow{2}{*}{settings}}&
				\multicolumn{3}{c|}{corresponding to the optimal quantum violation of the Bell inequality $\mathcal{B}_n$} \\
				\multicolumn{1}{|c|}{\multirow{2}{*}{($n$)}}&
				\multicolumn{3}{c|}{} \\
				\cline{2-4} 
				& $m=1$ & $m=2$ & $m=3$\\
				\hline \cline{1-4}
				&&&\\
				$2$  & $-\log_2 \qty[\frac{1}{4}\qty(1+\frac{1}{\sqrt{2}})] \approx 1.2284$ &$-\log_2 \qty[\frac{1}{4}\qty(1+\frac{1}{\sqrt{2}})] \approx 1.2284$& $-\log_2 \qty[\frac{1}{4}\qty(1+\frac{1}{\sqrt{2}})] \approx 1.2284$ \\
				&&&\\
				\hline
				&&&\\
				3  & $-\log_2 \qty[\frac{1}{4}\qty(1+\frac{1}{\sqrt{3}})] \approx 1.3425$ & $-\log_2 \qty[\frac{1}{4}\qty(1+\frac{1}{\sqrt{3}})] \approx 1.3425$ & $-\log_2 \qty[\frac{1}{4}\qty(1+\frac{1}{\sqrt{3}})] \approx 1.3425$ \\
				&&&\\
				\hline
				&&&\\
				4 & $-\log_2\qty[\frac{1}{12} \qty(3+\sqrt{6})] \approx 1.1388$ & $-\log_2 \qty[\frac{3}{8}] \approx 1.4150$  & $-\log_2 \qty[\frac{3}{8}] \approx 1.4150$ \\
				&&&\\
				\hline
				&&&\\
				5  & $-\log_2\qty[\frac{1}{4}\qty(1+\frac{\sqrt{2}+1}{\sqrt{2 \sqrt{2}+5}})]\approx 1.1025$ & $-\log_2\qty[\frac{1}{4}(1+\frac{1}{\sqrt{5}})]\approx 1.4667$& $-\log_2\qty[\frac{1}{4}(1+\frac{1}{\sqrt{5}})]\approx 1.4667$ \\
				&&&\\
				\hline
				&&&\\
				6  & $-\log_2\qty[\frac{1}{4}(1+\frac{3}{\sqrt{10}}]\approx1.0375$& $-\log_2\qty[\frac{1}{4}(1+\frac{1}{\sqrt{2}})]\approx 1.2284$ & $-\log_2\qty[\frac{1}{4}(1+\frac{1}{\sqrt{6}})]\approx1.5061$ \\
				&&&\\
				\hline \cline{1-4}
			\end{tabular}
			\caption{ The table shows the amount of certified randomness for different copies ($m\leq \lfloor n/2 \rfloor$) of bipartite maximally entangled state corresponding to different number of measurement settings $n \in\{1,2,3,4,5,6\}$. Clearly, it is seen that with the increasing number of $n$, the number of shared maximally entangled two-qubit state need to be increased to obtain a higher amount of certified randomness. The maximum amount of certified randomness for a particular $n$ can be achieved for $\lfloor n/2\rfloor$ copies of a maximally entangled two-qubit state which provide the optimal quantum violation of the Bell inequality $\mathcal{B}_{n}$.}
			\label{table1}
		\end{table*}
		
	\end{widetext}
	
	Finally, we evaluate the amount of certified randomness for arbitrary $n$-settings Bell inequality $(\mathcal{B}_n)$. The optimal quantum violation has been earlier derived to be $\langle\mathcal{B}_n\rangle_Q^{opt}=2^{n-1}\sqrt{n}$ \cite{Ghorai2018} and such optimal bound is achieved for $m=\lfloor \frac{n}{2} \rfloor$ copies of maximally entangled two-qubit state. The maximum joint probability, in this case, can be evaluated by following the simple argument. For this purpose, we recall the maximum success probability $(p_Q^{\ast}(b=x_y^{\delta}))$  of the communication game in \cite{Ghorai2018} given by $p_Q^{\ast}(b=x_y^{\delta})=\frac{1}{2}(1+\frac{1}{\sqrt{n}})$. Note that the probability $(p_Q^{\ast}(b=x_y^{\delta}))$ is the conditional probability of Bob's outcome $b$ corresponding to the measurements $B_{n,y}$ when Alice obtains the outcome $a$ after measuring $A_{n,i}$. Thus, using the Bayes rule of conditional probability, the maximum joint probability is given by
	\begin{eqnarray}\label{pmaxn}
		p^{*}(a,b|A_{n,1},B_{n,y},\rho_{AB}^{\otimes m})&=& p^{*}(a|A_{n,1},\rho_{AB}^{\otimes m}) \  p^{*}(b|a,A_{n,1},B_{n,y},\rho_{AB}^{\otimes m}) \nonumber\\
		&=&p^{*}(a|A_{n,1},\rho_{AB}^{\otimes m}) \ p_Q^{*}(b=x_y^{\delta}) \nonumber\\
		&=& \frac{1}{2} \ \times \ \frac{1}{2} \qty(1+\frac{1}{\sqrt{n}})
	\end{eqnarray} 
	
	Subsequently, the amount of certified randomness is then given by
	\begin{equation}
		\qty(R_{min})_n=2-\log_2 \qty[1+\frac{1}{\sqrt{n}}]
	\end{equation}
	
	It is interesting to note here that for large $n$, it is possible to certify close to 2 bits of $R_{min}$ using the Bell functional $\mathcal{B}_n$. It is important to remark here that close to two bits of DI guaranteed bound of randomness in terms of $R_{max}$ has already been shown by using different measurement contexts \cite{Acin2012, Law2014, Anderson2018, Woodhead2019}, our central motivation of this work is to show that the use of many copies of maximally entangled state in generating randomness quantified in terms of $R_{min}$ is more advantageous than the single copy.

	\section{Summary and outlook}\label{SecIV}
	
	In the present work, we investigate the possibility of certifying more randomness  quantified in terms of $R_{min}$from many copies of a maximally entangled two-qubit state than from a single copy. For this purpose, we first revisit the derivation of the quantum optimal value of the new family of Bell inequalities which was earlier introduced in the context of a RAC communication game and later, importantly, also demonstrated as a dimension witness. Next, by suitably quantifying the amount of randomness in terms of min.-Entropy, we evaluate the amount of randomness for different values of $n$. Specifically, we evaluate the amount of randomness corresponding to the optimal quantum violation of the invoked Bell inequality. Such evaluation of randomness only for the optimal quantum violation becomes sufficient for serving the purpose of our present work. 
	
	In particular, we explicitly show that the Bell inequality $\mathcal{B}_{n}$ for $n=2$ and $n=3$ case the guaranteed amount of randomness corresponding to the optimal quantum violation is obtained when a single copy of maximally entangled state is shared. Then, we show that the amount of randomness for the single copy of the maximally entangled state continues to decrease with the increase of $n\geq4$.  Next, for both the $n=4$ and $n=5$ cases, it is found that the amounts of certified randomness are increased if Alice and Bob share a pair of maximally entangled two-qubit states instead of a single copy. Such an amount of randomness again decreases for $n\geq6$. Moreover, for $n=6$, we obtain that the maximum amount of certified randomness is achieved when Alice and Bob share three pairs of maximally entangled two-qubit states. Hence, it follows from our demonstration that with the increasing number of measurement settings, the minimum amount of certified randomness depends on the number of maximally entangled two-qubit states shared between Alice and Bob.  Finally, we evaluate the maximum amount of $R_{min}$ corresponding to the optimal quantum value for an arbitrary number of measurement settings $n$. Such amount of randomness will be realized if Alice and Bob share $m=\lfloor n/2 \rfloor$ copies of maximally entangled two-qubit state. The findings of our results are illustrated in Table~\ref{table1} and Fig. \ref{figrn}.
	
	While our current study has limited advantage in randomness generation as compared to the existing protocols, such study may serve as a stepping stone towards the near future practical applications involving many copies of maximally entangled state. In particular, the present study may lead to the DI certification of an unbounded amount of randomness using the many copies of the maximally entangled state. Further, the present study then opens up an important avenue for the randomness expansion protocol in which a short and weak random string can be expanded into a long and strong certified provable secure randomness string. To date the randomness expansion protocols have been explored by using either the CHSH inequality or Hardy relations \cite{Ramanathan2018, Colbeck2012, Li2021, Liu2021, Bhavsar2021} which cannot certify more than a single copy of the bipartite maximally entangled state. Hence, it will be interesting to follow up the expansion as well as amplification protocols by invoking the new family of Bell inequalities $\mathcal{B}_n$ which has the capability of certifying more randomness from many copies of maximally entangled two-qubit states.

	\section*{Acknowledgments}
	We thank Souradeep Sasmal for his immense help in writing this paper. SSM acknowledges the UGC fellowship [Fellowship No. 16-9(June 2018)/2019(NET/CSIR)]. AKP acknowledges the support from the project DST/ICPS/QuST/Theme 1/2019/4. \linebreak
	
	\bibliography{references}

\end{document}